\documentclass[11pt]{article}
\usepackage{geometry}                
\usepackage{graphicx}
\usepackage{amssymb}
\usepackage{epstopdf}
\usepackage[T1]{fontenc}
\usepackage{color}
\usepackage{tabularx}
\usepackage{amsmath}
\usepackage{amsthm}

\newcommand{\bq}{\begin{eqnarray}}
\newcommand{\eq}{\end{eqnarray}}
\newcommand{\bitm}{\begin{itemize}}
\newcommand{\eitm}{\end{itemize}}

\def\R{\mathbb{R}}
\def\N{\mathbb{N}}

\def\Z {\mathbb{Z}}
\def\I{\large 1}
\def\lg{\langle }
\def\rg{\rangle }
\def\deq{\stackrel{\mathrm{def}}{=}}
\def\vpe{\varpi^{\sigma}}
\title{Action-angle coherent states for quantum systems with cylindric phase space}
\author{Isiaka Aremua, Jean Pierre Gazeau${}^{2}$\thanks{gazeau@apc.univ-paris7.fr}, and Mahouton Norbert Hounkonnou${}^{1}$\thanks{norbert.hounkonnou@cipma.uac.bj}\\
\\
${}^{1}$ International Chair of Mathematical Physics
and Applications \\
ICMPA-UNESCO Chair,
University of Abomey-Calavi\\
072 B.P. 50 Cotonou, Republic of Benin\\
${}^{2}$ Laboratoire APC, Univ Paris Diderot, Sorbonne Paris Cit\'e,  75205 Paris, France}

\begin{document}
\maketitle
\begin{abstract}
Quantum versions of cylindric phase space, like for the 
motion of a particle on the circle,  are  obtained through  different families of coherent states. The latter are  built from 
various probability distributions of the action variable. The method is illustrated with Gaussian distributions and 
uniform distributions on intervals, and resulting quantizations are explored. 
\end{abstract}

\section{Introduction}
\label{intro15}
In most of the introductory references in the literature devoted to quantum mechanics, the quantum versions of two simple models are presented, namely, 
the motion of a particle
on the circle and on an interval (e.g. the infinite square well potential).  
Experience with the harmonic oscillator suggests that the concept of coherent states \cite{klau85}
would be an important tool for the better understanding of the periodic motion of a quantum
particle. It is well known, essentially since Klauder \cite{klau63,klau95} and Berezin \cite{ber75}, (see also the stochastic quantum mechanics and quantum space-time programme based on the Prugovecki's views and comprehensively presented in \cite{ali85}), that 
one can easily achieve canonical quantization of the classical phase space by using standard coherent states. The coherent state quantization with 
its various generalizations reveals itself as an efficient tool for quantizing physical systems.
Recently,  this method has been implemented on various simple systems with   phase spaces like the complex plane with CS different of the standard ones \cite{cogagor10}, the cylinder \cite{main:ch15:gapie}, an infinite strip in the
plane \cite{main:ch15:gagaque, main:ch15:gagagitque,bergasiyou10}, or yet more exotic phase spaces like the finite set $\Z_d\times\Z_d$ \cite{cotgavou11} or paragrassmann algebras \cite{elgafreha10}.  
Starting from a solution to a version of the Stieltjes moment problem \cite{balgagi09A},  a family of coherent 
states has been also used to construct a  Fock-Bergmann representation  related to the particle quantization which  takes 
into account the circle topology of the classical motion. 
In \cite{balgagi09C},   canonical and  coherent state quantizations of a particle 
moving in a magnetic field  have been compared in the case of the non-commutative plane and semi-classical aspects have been explored.
Other examples are given in \cite{gazbook09} in which the method is explained at length and more complete references are given.

For the particular case   of a quantum
particle moving on a circle, i.e. when the phase space is topologically a cylinder,  coherent state constructions 
 have been   independently  proposed in \cite{main:ch5:debgo}, \cite{main:ch5:kopap}-\cite{main:ch5:hallmitch}. All these constructions lead to a certain type of coherent states: they are Gaussian in the sense that, as superpositions of angular momentum eigenstates $|n\rg$, $n\in \Z$ (and not $\N$!), their  Fourier coefficients involve Gaussian functions centered at  $n$.   It is noticeable that similar states were  proposed earlier by Chang  and Chi \cite{changshi86} for the treatment of a generalized quantum Chirikov map under the rational resonance condition $2\pi \hbar = M/N$ (see also \`Zyczkowski \cite{zyczkowski89}).

 The content of the present paper also concerns the motion of a quantum particle  on a circle, and more generally systems  for which the phase space  has cylindric topology.  Our  work lies in the continuation of those quoted  above and is also based on  developments elaborated in 
\cite{main:ch5:algahel,gaka11,bagagilev11}. We present  families of coherent states built from various probability distributions of the action 
variable for the motion on the circle. Our results  might reveal particularly relevant to recent models in superconducting circuit QED \cite{bovijoesdev98,naciu10} for which is raised again the longstanding question (see for instance \cite{carnieto68,busch01,galapon02}) of commutation relation between phase ($\sim$ angle) and number operator ($\sim$ angular momentum), more precisely excess Cooper pair number operator with spectrum $\Z$ and not just $\N$.  They could also offer new perspectives in the study of time behavior of quantum-chaotic phenomena.

The paper is organized as follows. In Section \ref{seccsquant} we give a short account of  the  coherent states quantization procedure with respect to a set $X$ of parameters equipped with a measure  
$\mu,$ and the statistical aspects leading to a Bayesian duality related to these states are summarized in Appendix \ref{statasp}. Section \ref{cylinder} is devoted to the construction of families of coherent states for the cylinder viewed as a phase space and  are associated with  various  distributions of the action (or angular momentum) variable. 
In Section \ref{quantcs}   the quantization of classical observables based on these various CS families is analyzed on a general level. 
In Section \ref{unif} the instructive although quite elementary case of  uniform distributions on intervals is worked out, and some 
semi-classical aspects are examined. Some further extensions of this work are investigated  in  Section \ref{conclu}.

\section{Coherent state quantization: the general setting}
\label{seccsquant}
Let $X$ be a set of parameters equipped with a measure $\mu$ and let $L^2(X, \mu)$ be its associated Hilbert space  of complex-valued square integrable functions with respect to $\mu$. Let us choose 
in $L^2(X, \mu)$ a finite or countable  orthonormal set $\mathcal{O}=\{\phi_n\, , \, n =\mathcal{F} \}$, with $\mathcal{F}$ some countable set ($\sim \N$ or $\sim \Z$ ...),
\begin{equation}\label{eqI1}
\lg \phi_m | \phi_n \rg = \int_{X}\overline{\phi_m(x)}\, \phi_n(x)\, \mu(dx) = \delta_{mn}\, ,
\end{equation}
 obeying the (crucial)  condition:
\begin{equation}\label{eqI2}
0< \sum_{n} \vert \phi_n (x)\vert^2 \deq \mathcal{N}(x) < \infty \,  \quad \mathrm{a.e.}\, . 
\end{equation}

Let $\mathcal{H}$ be a separable complex Hilbert space with orthonormal basis $\{|e_n\rg\, , \, n \in \mathcal{F} \}$, in one-to-one correspondence with the elements of  $\mathcal{O}$.   In particular,  it can be chosen as  the Hilbert subspace $\mathcal{K}_{\mathcal{O}}\deq \overline{\mathrm{span}(\mathcal{O})}$ in $L^2(X, \mu)$ itself.  We then define the  family of  states $\mathcal{F}_{\mathcal{H}}= \{|x\rg\, , \, x \in X \}$ in $\mathcal{H}$ as:
\begin{equation}\label{eqI3}
|x\rg = \frac{1}{\sqrt{\mathcal{N}(x)}}\sum_n \overline{\phi_n(x)}\, |e_n\rg\,.
\end{equation}
From Conditions (\ref{eqI1}) and (\ref{eqI2}) these ``coherent'' states are normalized, $\lg x| x\rg = 1$ and resolve the identity in $\mathcal{H}$:
\begin{equation}\label{eq14}
\int_X \mu(dx) \,\mathcal{N}(x) \, |x\rg \lg x | = \I_{{\mathcal H}}\, .
\end{equation}
The  relation (\ref{eq14}) allows us to implement a \emph{coherent state or frame quantization} of the set of parameters $X$ by associating to a function $X \ni x \mapsto f(x)$ that satisfies appropriate conditions the following operator in $\mathcal{H}$:
\begin{equation}\label{eqI5}
f(x) \mapsto A_f  \deq \int_X\mu(dx) \,\mathcal{N}(x) \, f(x)\, |x\rg \lg x |\, .
\end{equation}
The matrix elements of $A_f$ with respect to the basis $|e_n\rg$ are given by
\begin{equation}
\label{matelgen}
\left( A_f\right)_{nn'}= \lg e_n| A_f|e_{n'}\rg = \int_X\mu(dx) \, f(x) \, \overline{\phi_n(x)}\, \phi_{n'}(x)\, . 
\end{equation}
Operator $A_f$ is symmetric if $f(x)$ is real-valued,  bounded if $f(x)$ is bounded, self-adjoint if $f(x)$ real semi-bounded (through Friedrich's extension). 
In order to view  the ``upper'' symbol $f$ of $A_f$  as a quantizable object (with respect to the family $\mathcal{F}_{\mathcal{H}}$), a reasonable requirement   is that the so-called 
``lower symbol'' of $A_f$, 
defined as 
\begin{equation}
\label{lowsymgen}
\check{f} (x)\deq\lg x | A_f | x \rg = \int_X\mu(dx') \,\mathcal{N}(x') \, f(x')\, \vert\lg x|  x'\rg\vert^2 \, 
\end{equation}
 be a smooth function on $X$ with respect to some topology  assigned to the set $X$. 
  In Appendix \ref{statasp} we review some interesting statistical aspects of the above construction.

\section{The cylinder as a phase space for the motion on the circle}
\label{cylinder}
Quantization of the motion of a particle on the circle (like the quantization of polar coordinates in the plane) is an old question with so far 
mildly evasive answers. A large literature exists concerning the subject, more specifically devoted to the problem of angular localization and related 
Heisenberg inequalities \cite{carnieto68,busch01,galapon02}.
Let us apply our  scheme of coherent state quantization to this particular problem.  The observation set $X$ is  the phase space of a particle moving on the circle, precisely  the cylinder $X \equiv S^1 \times \R = \{ x\equiv (\varphi, J), \, | \, 0 \leq \varphi < 2\pi , \, J \in \R \}$, equipped with the measure $\mu(dx) = \frac{1}{2\pi}  \, dJ\, d\varphi $.

 We now introduce a probability distribution on the range of the  variable $J$. It is a non-negative, \underline{even}, well localized and normalized  
integrable  function 
\begin{equation}
\label{gprobdist}
\R\ni J \mapsto \varpi^{\sigma} (J)\,  ,\quad \varpi^{\sigma} (J)= \varpi^{\sigma} (-J)\,, \quad \int_{-\infty}^{+\infty} dJ\, 
\varpi^{\sigma} (J)=1\, , 
\end{equation}
where $\sigma >0$ is a kind of width parameter. 
This function must obey the following conditions:
\begin{itemize}
  \item[(i)] $0< \mathcal{N}^{\sigma}(J) \deq \sum_{n\in \Z}  \varpi^{\sigma}_n (J) < \infty$ for all $J\in \R$, where $\varpi^{\sigma}_n (J)\deq \varpi^{\sigma}_{0} (J-n)$,
  \item[(ii)] the Poisson summation formula is applicable to $\mathcal{N}^{\sigma}$:
  \begin{equation}
\label{poissvp}
\mathcal{N}^{\sigma}(J) = \sum_{n\in \Z}  \varpi^{\sigma}_n (J) = 
\sqrt{2\pi} \sum_{n\in \Z}  e^{-2\pi i nJ}\widehat{\varpi}^{\sigma}_{n}(2\pi n)\, ,  
\end{equation}
where $\hat{\varpi}^{\sigma}$ is the Fourier transform of $\vpe$,
  \item[(iii)]  its limit  at $\sigma \to 0$, in a distributional  sense, is the Dirac distribution:
  \begin{equation}
\label{varpidi}
\varpi^{\sigma} (J) \underset{\sigma \to 0}{\to} \delta(J)\,,
\end{equation} 
 \item[(iv)]  the limit  at $\sigma \to \infty$ of its Fourier transform is proportional to the characteristic function of the singleton $\{0\}$:
  \begin{equation}
\label{varpidir}
\widehat{\varpi}^{\sigma} (k) \underset{\sigma \to \infty}{\to} \frac{1}{\sqrt{2\pi}}\, \delta_{k0}\,,
\end{equation} 
\item[(v)] considering the \emph{overlap matrix} of the two distributions $J \mapsto \vpe_n(J)$, $J \mapsto \vpe_{n'}(J)$ with matrix elements,
 \begin{equation}
\label{varpcor}
\vpe_{n,n'} = \int_{-\infty}^{+\infty} dJ\, \sqrt{\vpe_n(J)\, \vpe_{n'}(J)} \leq 1\, , 
\end{equation}
we impose the two conditions 
\begin{align}
\label{cond1}
    \vpe_{n,n'} &\to 0 \quad \mbox{as} \quad  n-n' \to \infty  \quad \mbox{at fixed}\ \sigma\, , \\
 \label{cond2}  \exists\, N_0\geq 1& \quad \mbox{such that} \quad   \vpe_{n,n'} \underset{\sigma \to \infty}{\to} 1 \quad   \ \forall\, n, \, n'\, \ \mbox{such that} \ \vert n-n'\vert \leq N_0 \, .
\end{align} 
 \end{itemize} 
  Properties (ii) and (iv) entail that $ \mathcal{N}^{\sigma}(J)  \underset{\sigma \to \infty}{\to} 1$. Also note the properties of the overlap matrix elements $\vpe_{n,n'}$ due to the properties of $\vpe$:
\begin{equation}
\label{propcorr}
\vpe_{n,n'}= \vpe_{n',n}=\vpe_{0,n'-n}= \vpe_{-n,-n'}\,, \quad \vpe_{n,n} = 1\, \quad \forall\, n, n' \in \Z\, .
\end{equation}
The most immediate (and historical) choice for $\varpi^{\sigma} (J)$  is  Gaussian, i.e. $\varpi^{\sigma} (J)= \sqrt{\frac{1}{2\pi\sigma^2}}\,e^{-\frac{1}{2\sigma^2} J^2 }$ (for which the $N_0$ in (\ref{cond2}) is $\infty$),  as it appears under various forms in  
 the existing literature on the subject \cite{main:ch5:debgo}--\cite{main:ch5:hallmitch}. In Appendix \ref{normalaw} we recall a few features of CS issued from such a choice. 

Let us now introduce the weighted Fourier exponentials:
\begin{equation}
\label{circphin1}
\phi_n (x) = \sqrt{\varpi^{\sigma}_n(J)}  \,e^{  i n\varphi}\, , \quad n\in \Z\,.
\end{equation}
These functions form the countable  orthonormal system in $L^2(X,\mu(dx))$ needed to construct coherent states in agreement with the procedure explained in Section \ref{seccsquant}. In consequence, the  correspondent family of coherent states on the circle reads as:
\begin{equation}
\label{ccs}
| J, \varphi \rangle =  \frac{1}{\sqrt{{\mathcal N}^{\sigma} (J)}} \sum_{n \in \Z} \sqrt{\varpi^{\sigma}_n(J)}  \,e^{- i n\varphi} | e_n\rangle\, .
\end{equation}

As expected, these states  are normalized and resolve the unity. They overlap as:
\begin{equation}
\label{overlap}
\lg J,\varphi|J',\varphi'\rg =  \frac{1}{\sqrt{{\mathcal N}^{\sigma} (J)\, {\mathcal N}^{\sigma} (J')}}\sum _{n \in \Z} \sqrt{\varpi^{\sigma}_n(J)\, \varpi^{\sigma}_n(J')}  \,e^{- i n(\varphi-\varphi')}\, . 
\end{equation}

As explained in Appendix \ref{statasp},  the function $\varpi^{\sigma} (J)$ gives rise to a double probabilistic interpretation \cite{main:ch5:algahel,gazbook09}:
\begin{itemize}
  \item For all $J$ viewed as a shape parameter, there is the discrete distribution, 
 \begin{equation}
\label{discrprobC}
\Z \ni n \mapsto \vert \lg e_n|x\rg\vert^2=  \frac{ \varpi^{\sigma}_n (J)}{{\cal N}^{\sigma} (J)} \, .
\end{equation}
This probability, of genuine quantum nature,   concerns experiments performed on the system described by the Hilbert space $\mathcal{H}$ within some experimental protocol, say $\mathcal{E}$, in order to measure the  spectral values of the  self-adjoint operator  acting in $\mathcal{H}$ and having the discrete spectral resolution $ \sum_n a_n |e_n\rg\lg e_n|$. For $a_n=n$ this operator is  the quantum angular momentum, as will be shown in the next section.

 \item For each $n$, there is the continuous distribution on the cylinder $X$ (reps. on $\R$) equipped with its  measure $dJ\, d\varphi/2\pi$ (resp. $dJ$), 
\begin{equation}
\label{contprobC}
X \ni (J, \varphi) \mapsto \vert \phi_n (J,\varphi) \vert^2 = \varpi^{\sigma}_n (J)\quad (\mbox{resp.}\ \quad \R \ni J \mapsto \varpi^{\sigma}_n (J))\,.
\end{equation}
This probability, of classical nature and uniform on the circle,  determines the CS quantization of functions of $J$, as will be seen in the next section.
 \end{itemize}
 
\section{Quantization of classical observables with  CS on the circle}
\label{quantcs}
\subsection{General setting}
By virtue of the CS quantization scheme  described in Section \ref{seccsquant}, the quantum operator (acting on ${\mathcal
H}$) associated with the classical observable $f(x)$ is obtained through
\begin{equation}
A_f := \int_X f(x) | x\rangle \langle x| \,  \mathcal{N}(x) \mu(dx) = \sum_{n,n'} \left(A_f\right)_{nn'} \, |e_n\rg \lg e_{n'}|\, ,
\label{oper15}
\end{equation}
where
\begin{equation}
\label{matelAf15}
\left(A_f\right)_{nn'} = \int_{-\infty}^{+\infty}dJ\, \sqrt{\vpe_n(J)\, \vpe_{n'}(J)}\,\frac{1}{2 \pi}\int_0^{2 \pi}d \varphi\, e^{-i(n-n')\varphi}\, f(J,\varphi)\, .
\end{equation}
The lower symbol of $f$ is  given by:
\begin{equation}
\label{lowsymbvp}
\check{f} (J,\varphi) = \lg J,\phi | A_f | J,\phi \rg 
= \int_{-\infty}^{+\infty}dJ' \int_{0}^{2\pi}
\frac{d\varphi'}{2\pi}\,\mathcal{N}^{\sigma}(J') \, f(J',\varphi')\, \vert\lg J,\phi|  J',\varphi'\rg\vert^2 \, .
\end{equation}
If $f$ is depends on $J$  only, $f(x) \equiv f(J)$, then $A_f$ is diagonal with matrix elements that are $\vpe$ transforms of $f(J)$:

\begin{equation}
\label{diagmatelAf15}
\left(A_{f(J)}\right)_{nn'} = \delta_{nn'}\int_{-\infty}^{+\infty}dJ\, \vpe_n(J)\, f(J)= \delta_{nn'} \lg f\rg_{\vpe_n}\, ,
\end{equation}

where $ \lg \cdot \rg_{\vpe_n}$ designates the mean value w.r.t. the distribution $J\mapsto \vpe_n(J)$. 
 For the most basic case, $f(J) = J$,    our assumptions  on $\vpe$ give
\begin{equation}  A_J  = \int_{X} \mu (dx)  \mathcal{N}^{\sigma}(J)\, J\, | J,\varphi \rangle \langle J, \varphi |  = \sum_{n \in \Z}
n\, | e_n\rangle \langle e_n| \, .
\label{Jsym}
\end{equation}
This is nothing  but the angular momentum operator (in unit $\hbar = 1$), which reads  $A_J = -i\partial/\partial \theta$ in angular position representation, i.e. when $\mathcal{H}$ is chosen as $L^2(S^1,d\theta/2\pi)$ with orthonormal basis $|e_n\rg \equiv e^{in\theta}$ (Fourier series). The covariance property of the coherent states with respect to  rotations is a direct consequence of (\ref{diagmatelAf15}):
\begin{equation}
\label{covCS}
e^{i \theta A_J}|J,\varphi\rg = |J, \varphi - \theta\rg\, . 
\end{equation}
The quantization of $f(J) = J^2$, i.e. the kinetic energy of the particle in suitable units, produces a quantum spectrum which behaves like $n^2$:
\begin{equation}  A_{J^2}  = \int_{X} \mu (dx)  \mathcal{N}^{\sigma}(J)\, J\, | J,\varphi \rangle \langle J, \varphi |  = c\, \I_{{\mathcal H}} + \sum_{n \in \Z}
n^2\, | e_n\rangle \langle e_n|\, ,
\label{envp}
\end{equation}
where 
\begin{equation}
\label{avvp}
c = \int_{-\infty}^{+\infty} dJ\, J^2\, \vpe(J) = \lg J^2 \rg_{\vpe_n}\,.
\end{equation}
 We can understand through 
this quantization procedure the probabilistic origin of the elementary quantum of energy, i.e. the difference between the classical 
zero energy point and  the quantum ``vacuum energy''. 

If $f$ depends on $\varphi$ only,   $f(x) \equiv f(\varphi)$, we have
\begin{align}  A_{f(\varphi)} = & \int_{X} \mu (dx)  \mathcal{N}^{\sigma}(J) f(\varphi) \,  | J,\varphi \rangle \langle J, \varphi |  \\
&= \sum_{n,n' \in \Z}
\vpe_{n,n'} \,c_{n-n'}(f)| e_n\rangle \langle e_{n'} |\,  ,
\label{f(beta)15}
\end{align}
 where 
 $c_{n}(f)$ is the $n$th Fourier coefficient of $f$. At a 
first look at (\ref{f(beta)15}), one  understands that the more  distributions overlap, the more  the non commutativity is enhanced.  
In particular, we have 
 \begin{itemize}
\item the self-adjoint ``angle'' operator corresponding to the $2\pi$-periodic saw function $B(\varphi)$ defined by periodic extension of $B(\varphi) = \varphi$  for $0\leq \varphi < 2\pi$, and abusively denoted in the sequel by $\varphi$,
\begin{equation}
\label{opangle15}
 A_{\varphi} = \pi \I_{{\mathcal H}} + i\, \sum_{n\neq n'}\frac{\vpe_{n,n'}}{n-n'}\,| e_n\rangle \langle e_{n'} |\, ,
 \end{equation}
\item the operator Fourier fundamental  harmonics corresponding to elementary Fourier exponential,
\begin{equation}
\label{opfourier15}
A_{e^{\pm i\varphi}} = \, \vpe_{1,0}\sum_{n}
 | e_{n \pm 1}\rangle \langle e_n |\, , \quad A_{e^{\pm i\varphi}}^{\dag}= A_{e^{\mp i\varphi}}\, . 
\end{equation}
 \end{itemize}
We remark that $A_{e^{\pm i\varphi}}\, A_{e^{\pm i\varphi}}^{\dag}=  A_{e^{\pm i\varphi}}^{\dag}\,A_{e^{\pm i\varphi}}= (\vpe_{1,0})^2 1_d$. Therefore this operator fails to be unitary. It is ``almost'' unitary at large $\sigma$ since the factor
$(\vpe_{1,0})^2$ can  be made  arbitrarily close to 1 at large $\sigma$ as a consequence of Requirement (\ref{cond2}). In the Fourier series realization of ${\mathcal H}$, for which the kets $ | e_n \rangle$
are the Fourier exponentials $e^{i\,n\theta}$,  the operators $A_{e^{\pm i\varphi}} $ are multiplication operator by  $e^{\pm i\theta}$ up to the factor
$\vpe_{1,0}$. 

\subsection{Elementary commutators and classical limit}
The  commutation rules 
\begin{equation}
\label{cancomrulcyl15}
\lbrack A_J, A_{e^{\pm i\varphi}} \rbrack = \pm A_{ e^{\pm i\varphi}}
\end{equation}
are  canonical in the sense that they are  in exact correspondence with the classical Poisson brackets 
\begin{equation}
\label{poisbraccyl15}
\left\{ J, e^{\pm i\varphi} \right\} = \pm i e^{\pm i\varphi}\, .
\end{equation}
(For other  non trivial commutators having this exact correspondence in the Gaussian case, see \cite{main:ch15:rabhure}.)
In consequence, our CS quantization based on a choice of $\vpe$ fulfilling Conditions (i)-(v) respects the underlying symmetry $SO(2)\ltimes \R^2$  of the cylinder viewed as a phase space. Indeed, after introducing a positive constant $\lambda$,  we deduce from (\ref{cancomrulcyl15}) the commutation rules 
\begin{equation}
\label{creucl1}
\lbrack A_J, A_{\lambda \cos\varphi} \rbrack =  iA_{\lambda \sin\varphi}\, , \quad \lbrack A_J, A_{\lambda \sin\varphi} \rbrack = -iA_{\lambda \cos\varphi}\,, \quad \lbrack A_{\lambda \cos\varphi}, A_{\lambda \sin\varphi} \rbrack = 0\, . 
\end{equation}
They are those verified by generators of a unitary representation of the Euclidean group of the plane. 

One could be puzzled by  commutators of the type:
\begin{equation}
\label{othcomcyl15}
\lbrack A_J, A_{f(\varphi)} \rbrack =  \sum_{n, n'}(n-n')\,  \vpe_{n,n'}
\,c_{n-n'}(f)\, | e_n\rangle \langle e_{n'} |\, ,
\end{equation}
and, in particular, for the angle operator itself:
\begin{equation}
\label{ccrcir15}
 \lbrack A_J, A_{\varphi} \rbrack = i \sum_{n \neq n'}
 \vpe_{n,n'}\, | e_n\rangle \langle e_{n' } |\, ,
 \end{equation} 
to be compared with the classical $\left\{ J, \varphi \right\} = 1$. One observes that the overlap matrix completely encodes  the basic commutator between quantized canonical variables action-angle. 

Because of  the required properties of the distribution $\vpe$ the departure of the r.h.s. of Eq. (\ref{ccrcir15}) from the canonical r.h.s. $-i 1_{\mathcal{H}}$ can be bypassed by examining the behavior of the lower symbols at large $\sigma$. For an original function depending on $\varphi$ only we have the Fourier series
\begin{equation}
\label{lowsymbfphi}
\check{f}(J_0, \varphi_0)=  \langle J_0, \varphi_0 | A_{f(\varphi)} |  J_0, \varphi_0 \rangle    =  c_0(f) + \sum_{m \neq 0} d_m^{\sigma}(J_0)\,  \vpe_{0,m}\, c_m(f)\, e^{ i m \varphi_0}\, , 
\end{equation}
with 
\begin{equation}
\label{coeffdfphi}
d_m^{\sigma}(J)= \frac{1}{\mathcal{N}^{\sigma}(J)}\sum_{r=-\infty}^{+\infty} \sqrt{\vpe_{r}(J)\vpe_{m+r}(J)} \leq 1\, ,
\end{equation}
the last inequality resulting from Condition (i) and Cauchy-Schwarz inequality. If we further  impose the condition that $d_m^{\sigma}(J) \to 1$ uniformly as $\sigma \to \infty$, then  the lower symbol $\check{f}(J_0, \varphi_0)$ tends to the Fourier series of the original function $f(\varphi)$. A similar result is obtained for the lower symbol of the commutator (\ref{othcomcyl15}):
\begin{equation}
\label{lowsymcom}
 \langle J_0, \varphi_0 | \lbrack A_J, A_{f(\varphi)} \rbrack |  J_0, \varphi_0 \rangle    =  \sum_{m \neq 0} d_m^{\sigma}(J_0)\,  \vpe_{0,m}\, m\,c_m(f)\, e^{ i m \varphi_0}\, , 
\end{equation}
and in particular,

\begin{equation}
\label{lowsymcomangle}
 \langle J_0, \varphi_0 | \lbrack A_J, A_{\varphi} \rbrack |  J_0, \varphi_0 \rangle    =  
i \sum_{m \neq 0} d_m^{\sigma}(J_0)\,  \vpe_{0,m}\, e^{ i m \varphi_0}\, . 
\end{equation}

Therefore, with the condition that $d_m^{\sigma}(J) \to 1$ uniformly as $\sigma \to \infty$, we obtain at this limit  
\begin{equation}
\label{lowsymcomlim}
 \langle J_0, \varphi_0 | \lbrack A_J, A_{\varphi} \rbrack |  J_0, \varphi_0 \rangle \underset{\sigma \to \infty}{\to}   -i +  i\sum_{m } \delta(\varphi_0 - 2 \pi m)\, . 
\end{equation}
So we asymptotically (almost) recover the classical canonical commutation rule except for the singularity at the origin $\mathrm{mod}\, 2\pi$, a logical consequence of the discontinuities of the saw function $B(\varphi)$ at these points.

\subsection{Other semi-classical aspects}
We have  tested in the previous section a few semi-classical features of the coherent states (\ref{eqI3})  by studying how lower symbols of operators $A_f$
approach the original  $f(J,\varphi)$.  When the latter is semi-bounded from below, another possible  test consists in evaluating, at large $\sigma$,  the relative error function $\mathrm{rerr}_C(J,\varphi;f)$:
\begin{equation}
\mathrm{rerr}_C(J,\varphi;f) \deq \left\vert \frac{\langle J,\varphi| A_f| J,\varphi\rangle- f(J,\varphi)}{f(J,\varphi)+ C}\right\vert\,,
\label{error}%
\end{equation}
where the constant $C$ has to be added to $f$ in order that the denominator does not cancel. Of course, $C$ should not be chosen too large.  It is also possible to avoid  such a precaution by restricting the study of this function to the positive part of the range of $f$.

Another interesting exploration is the temporal behavior of the lower symbol, given some classical Hamiltonian function $H(J,\varphi)$  and its quantum version $A_H$,  once  the initial condition $(J_0,\varphi_0)$ has been chosen in the phase space. Thus, we can explore analytically and numerically expressions of the type
\begin{align}
\label{lsevopJ}
&\langle J_0,\varphi_0| e^{-i A_H t }A_J  e^{i A_H t}| J_0,\varphi_0\rangle \\
\label{lsevopth}&\langle J_0,\varphi_0| e^{-i A_H t} A_{\varphi}  e^{i A_H t}| J_0,\varphi_0\rangle \,.
\end{align}

Moreover,  the resolution of the identity (\ref{eq14}) allows for a probability distribution 
$(J,\vpe) \mapsto \mathcal{N}^{\sigma}(J)\,\vert \lg J,\varphi|J_0, \varphi_0\rg\vert^2 \equiv \rho_{|J_0, \varphi_0\rg}(J,\varphi)$ 
on the cylindric phase space. It is natural to consider this distribution as a localization   measure  on the phase space. Hence, 
given a Hamiltonian $H(J,\varphi)$, it is also natural to explore analytically and numerically the time evolution of such a distribution, 
\begin{equation}
\label{tevdist}
t\mapsto \rho_{e^{-i A_H t}|J_0, \varphi_0\rg}(J,\varphi)= \mathcal{N}^{\sigma}(J)\vert \lg J,\varphi|e^{-i A_H t}|J_0, \varphi_0\rg\vert^2\, ,
\end{equation}
and to compare it with the classical phase trajectory on the cylinder. 
For instance, in the Gaussian case, and with the Hamiltonian $H=J^2$ of the free motion on the circle,  
we can study the time evolution of the following series obtained either 
from Eq. (\ref{overlapgcs1}) or Eq. (\ref{overlapgcs2}) in Appendix \ref{normalaw}:
\begin{align}
\label{tevdist1}
 \rho_{e^{-i A_H t}|J_0, \varphi_0\rg}(J,\varphi)&= \frac{e^{-\frac{1}{4\sigma^2}(J-J_0)^2}}{ 2\pi \sigma^2\,  \mathcal{N}^{\sigma}(J_0)}\left\vert \sum_{n\in \Z} e^{-\frac{1}{2\sigma^2}(\frac{J+J_0}{2}- n)^2}\, e^{i  (n(\varphi-\varphi_0) -n^2 t)}\right\vert^2\\
 &=  \frac{e^{-\frac{1}{4\sigma^2}(J-J_0)^2}}{ \mathcal{N}^{\sigma}(J_0)}\left\vert \sum_{n\in \Z} e^{-\frac{\sigma^2}{2}(\varphi-\varphi_0-2\pi n)^2}\, e^{-i (\pi n(J+J_0) + n^2 t)}\right\vert^2\, . 
 \label{tevdist2}
 \end{align}
 At small $\sigma$, we deduce from (\ref{tevdist1}) that this distribution concentrates at $J=J_0$ under the condition that $J_0 \in \Z$, whatever the  
values of $t$ 
and $\varphi -\varphi_0$. At large $\sigma$, we deduce from (\ref{tevdist2}) that it vanishes if $\varphi -\varphi_0 \notin 2\pi \Z$ and goes to 
1 otherwise,  whatever the values of $J$, $J_0$ and $t$. We could conclude from such a superficial analysis  that such coherent states have no correct 
semi-classical phase space aspect in terms of time evolution. Actually, the semi-classical analysis is made more consistent by reintroducing into the 
above expressions physical quantities involving the Planck constant $\hbar$  and to examine their behavior at large $J_0$ (more details are given in 
\cite{gaka11}).

\section{CS Quantization of the motion on the circle with uniform distribution on intervals }
\label{unif}
We now apply the above material to the  (apparently trivial) case of a uniform distribution on an interval centered at the origin and with length $2 \sigma$:
\begin{equation}
\label{distint}
\vpe(J) = \frac{1}{2\sigma}\chi_{[-\sigma, \sigma]}(J)\,.
\end{equation}
A first limitation on the range of $\sigma$ is necessary because of Condition (i):
\begin{equation}
\label{limsig1}
 \sigma \geq \frac{1}{2}\, .
\end{equation}
Indeed, the normalization function reads in the present case
\begin{equation}
\label{normun}
\mathcal{N}^{\sigma}(J) = \frac{1}{2\sigma}\left\lbrack 1 + \sum_{n\in \Z} \chi_{[n +1 -\sigma, n+ \sigma]}(J)  \right\rbrack\, , 
\end{equation}
and would vanish for all $J \in  \cup_{n \in \Z}(n+\sigma, n+1-\sigma)$ if $\sigma < 1/2$. 
This periodic crenel function, period 1,  assumes only two values, $1/(2\sigma)$ and $1/\sigma$. In particular $\mathcal{N}^{\sigma}(n) = 1/(2\sigma)$ and $\mathcal{N}^{\sigma}(n+1/2) = 1/\sigma$ for all $n\in \Z$. For $\sigma = 1/2$ or $\sigma = 1$ it is 
equal a.e. to 1.

As is well known, the  Fourier transform of $\vpe(J)$ is a  cardinal sine,
\begin{equation}
\label{fourun}
\widehat{\vpe}(k) = \frac{1}{\sqrt{2\pi}}\, \sin_{\mathrm{c}}(\sigma k)\, , 
\end{equation}
which is at $k=0$  equal to $1/\sqrt{2\pi}$ for any $\sigma$. 
It is then clear that  conditions (i)-(iv) are fulfilled. 

Let us impose the supplementary limitation in the choice of $\sigma$,
\begin{equation}
\label{limsig2}
 \sigma \leq 1\, .
\end{equation}
Such a choice prevents us to examine limits at large  $\sigma$. On the other hand it makes analytic computations easer. 
Thus the following equation
\begin{align}
\label{produnif}
\nonumber\sqrt{\vpe_n(J)\, \vpe_{n'}(J)}&= \frac{1}{2\sigma}\left\lbrack\delta_{n \,n'}\chi_{[n  -\sigma, n+ \sigma]}(J) + \right.\\
&+\left. \delta_{n \,n'+1}\, \chi_{[n  -\sigma, n-1 +\sigma]}(J) + \delta_{n \,n'-1}\, \chi_{[n +1 -\sigma, n+\sigma]}(J)\right\rbrack\, , 
\end{align}
shows  that only nearest neighbors overlap:
\begin{equation}
\label{corun}
\vpe_{n,n'}=\delta_{n \,n'} + \left(1-\frac{1}{2\sigma}\right)\,[ \delta_{n \,n'+1} + \delta_{n \,n'-1}]\, , 
\end{equation}  
which simply means that $\vpe_{1,0}= 1-1/(2\sigma)$ and $\vpe_{n,n'} = 0$ for all $n$, $n'$, such that $\vert n-n'\vert > 1$. Note that the overlap is lost at the lowest limit $\sigma = 1/2$. Increasing the lowest upper bound in (\ref{limsig2}) would enlarge the overlap.

The quantization of \underline{any} locally integrable function $f(J,\varphi)$ produces
a tridiagonal matrix $A_f$ (which is Jacobi if $f$ is real):
\begin{align}
\label{quantunif}
\nonumber \left(A_f\right)_{nn'} &= \frac{1}{2\sigma}\left\lbrack\delta_{n\,n'}\int_{n-\sigma}^{n+\sigma}dJ\, \,\frac{1}{2 \pi}\int_0^{2 \pi}d \varphi\, f(J,\varphi) + \right.\\
\nonumber &+\delta_{n\, n'+1}\int_{n- \sigma}^{n -1 +\sigma}dJ\,\frac{1}{2 \pi}\int_0^{2 \pi}d \varphi\, e^{-i\varphi}\, f(J,\varphi)\\
&\left.+ \delta_{n\,  n'-1} \int_{n+ 1 -\sigma}^{n+\sigma}dJ\,\frac{1}{2 \pi}\int_0^{2 \pi}d \varphi\, e^{i\varphi}\, f(J,\varphi)\right\rbrack\, .
\end{align}

\subsection{Quantization of elementary observables}

We now specify the procedure to the most elementary functions $f(J,\varphi)$,
 noting that this specification from the general case  is straightforward.
\begin{enumerate}
\item[(i)]{Angular momentum or action operator}
 
It is given by the expression
\begin{equation}  A_J  = \int_{X} \mu (dx)  \mathcal{N}^{\sigma}(J)\, J\, | J,\varphi \rangle \langle J, \varphi |  = \sum_{n \in \Z}
n\, | e_n\rangle \langle e_n| ,
\label{Jsymun}
\end{equation}
which  coincides with (\ref{Jsym}).
\item[(ii)]{Energy operator} 

The energy operator is expressed by
    \begin{equation}  A_{J^2}  = \int_{X} \mu (dx)  \mathcal{N}^{\sigma}(J)\, J^{2}\, | J,\varphi \rangle \langle J, \varphi |  
= \frac{\sigma^2}{3}\, \I_{{\mathcal H}} + \sum_{n \in \Z}
n^2\, | e_n\rangle \langle e_n|\, 
\label{enrgun}
\end{equation}
  as it should from the relation (\ref{envp}) defined for an arbitrary distribution. The first right hand side term $\sigma^2/3$ indeed represents
the average of the classical energy with respect to the uniform probability distribution on the interval $[n-\sigma, n+\sigma],\,n \in \Z$.
  \item[(iii)]{Elementary Fourier harmonic operator}

     The operator ``Fourier fundamental  harmonics'' is defined by
\begin{equation}
\label{opfourierun}
A_{e^{\pm i\varphi}} = \left(1-\frac{1}{2\sigma}\right)\sum_{n \in \Z}
 | e_{n \pm 1}\rangle \langle e_n |\, .
\end{equation}
It  becomes null operator at the lowest limit $\sigma = 1/2$ whereas it  is one-half of the expected one at the  upper limit $\sigma = 1$. 
\item[(iv)]{Angle operator}

The angle operator is provided by  

\begin{equation}
\label{opangleun}
A_{\varphi} = \pi 1_{\mathcal H} + i \left(1-\frac{1}{2\sigma}\right)\sum_{n \in \Z}\left[|e_{n}\rangle \langle e_{n-1}| 
- |e_{n}\rangle \langle e_{n+1}|  \right] = A_{\pi - 2\sin{\varphi}} \, .
 \end{equation}

Hence its amounts to replace the angle function by the  two first terms of its Fourier series,
\begin{equation}
\label{angfour}
B(\varphi) = \pi - 2 \sum_{n \geq 1} \frac{\sin(n \varphi)}{n}\, .
\end{equation}
Note that the same operator can correspond to more than one classical observable and that for $\sigma = 1/2$ the angle operator  reduces to the classical angle average, namely $\pi$. 
 \end{enumerate}

 \subsection{Some commutators}
 \begin{enumerate}
  \item[(i)]{Commutator of action and angle operators}
 
 The evaluation of the commutator of the  action and angle operators is just proportional to the ``free'' infinite tridiagonal Jacobi matrix \cite{kisi03}:
  
\begin{equation} 
\label{ccrcirun}
[A_{J}, A_{\varphi}] = i \left(1-\frac{1}{2\sigma}\right)\sum_{n \in \Z}\left[|e_{n}\rangle \langle e_{n-1}| 
+ |e_{n}\rangle \langle e_{n+1}|  \right] = i A_{2\cos{\varphi}}\, ,
 \end{equation} 

  an expression which is consistent with (\ref{opangleun}) and the fact that $A_J$ acts as $-i\partial/\partial \varphi$ in the space 
$L^2(S^1, d\varphi/2\pi)$. This expression has to be compared with the classical Poisson bracket
$\{J, \varphi\} = 1$. It is well known that the spectral measure of the Jacobi matrix 
\begin{equation*}
\begin{pmatrix}
  0    & 1  &0 & 0 &  \cdots \\
   1   & 0 & 1 &0  &  \cdots\\
   0  & 1 & 0 &1  &  \cdots\\
    0   & 0 & 1 &0  &  \cdots\\ 
\end{pmatrix}
\end{equation*}
 is supported by $[-2,2]$, and so the spectrum of the commutator (\ref{ccrcirun}) is continuous and equal to the  interval $i\,[-2 -1/\sigma, 2-1/\sigma]$. 
 \item[(ii)]{Commutator of energy-angle operators}
 
 Similarly, we get for the commutator of energy-angle operators:
  
\begin{eqnarray} 
[A_{J^{2}}, A_{\varphi}] &=&  i \left(1-\frac{1}{2\sigma}\right)\sum_{n \in \Z}\left[(2n-1)|e_{n}\rangle \langle e_{n-1}| + (2n+1)
|e_{n}\rangle \langle e_{n+1}|\right] \cr
&=& 2i A_{J} A_{2\cos{\varphi}} + A_{2\sin{\varphi}} \, ,
 \end{eqnarray} 

an expression to be compared with the classical Poisson bracket
$\{J^2, \varphi\} = 2J$.

 \item[(iii)]{Commutator of action-elementary harmonics operators}
 
We have
\begin{equation}{\label{eq00}}
[A_{J}, A_{e^{\pm i\varphi}}] = \pm\left(1-\frac{1}{2\sigma}\right)\sum_{n\in \Z} |e_{n\pm1}\rangle \langle e_{n}| = \pm A_{e^{\pm i\varphi}}\, , 
\end{equation}
to be compared with $\{J, e^{\pm i\varphi}\} = \pm i e^{\pm i\varphi}.$
 \item[(iv)]{Commutator of energy-elementary harmonics operators}
\begin{equation}{\label{eq01}}
[A_{J^{2}}, A_{e^{\pm i\varphi}}] 
= A_{e^{\pm i\varphi}} \pm 2\left(1-\frac{1}{2\sigma}\right)\sum_{n \in \Z}n|e_{n\pm 1}\rangle \langle e_{n}|\, , 
\end{equation}
to be compared with  $\{J^{2}, e^{\pm i\varphi}\}  =  \pm i\, 2 J e^{\pm i\varphi}.$
 \end{enumerate}

 \subsection{Some lower symbols}
 The mean values of the above described operators with respect to the  coherent states 
for the  distribution  $\varpi^{\sigma}$ are obtained as linear superposition of 
distributions $\varpi^{\sigma}_{n}, n \in \Z$ as follows. 

 \begin{enumerate}

\item[(i)]{Angular momentum or action} 

The lower symbol of the angular momentum operator $A_{J}$ is given by
\begin{equation}
\langle J_{0}, \varphi_{0}|A_{J}|J_{0}, \varphi_{0}\rangle = 
\frac{1}{\mathcal N^{\sigma}(J_{0})}
\sum_{n\in \Z}n \, \varpi^{\sigma}_{n}(J_{0})= 
\frac{1}{\mathcal N^{\sigma}(J_{0})}\sum_{n\in \Z}\frac{n}{2\sigma}\; \,\chi_{[n-\sigma, n+\sigma]}(J_{0})\, ,
\end{equation}

\item[(ii)]{Energy}

\begin{equation}
 \langle J_{0}, \varphi_{0}|A_{J^{2}}|J_{0}, \varphi_{0}\rangle = \frac{\sigma^{2}}{3} + 
\frac{1}{\mathcal N^{\sigma}(J_{0})}\sum_{n\in \Z}\frac{n^{2}}{2\sigma}\; \,\chi_{[n-\sigma, n+\sigma]}(J_{0}).
\end{equation}
In the case of non correlation  $\sigma = 1/2$, i.e. in the commutative situation 
 the intervals $[n-\sigma, n+\sigma]$ are those separating successive 
half-integers $\{n+1/2\},$   $n \in \Z$. Then, if $J_{0}$ is a half integer, there exists an integer $n_{0} \in \Z$ such that 
$\chi_{\{n_{0}+1/2\}}(J_{0}) = 1$ and $\chi_{\{n+1/2\}}(J_{0}) = 0$ for $n \neq n_{0}$.  Therefore, we arrive at
\begin{align}
&\langle J_{0}, \varphi_{0}|A_{J}|J_{0}, \varphi_{0}\rangle =
\frac{1}{\mathcal N^{\sigma=1/2}(J_{0})}n_{0}\, , \\ 
&\langle J_{0}, \varphi_{0}|A_{J^{2}}|J_{0}, \varphi_{0}\rangle = \frac{1}{12} + \frac{1}{\mathcal N^{\sigma=1/2}(J_{0})}n^{2}_{0}\,.
\end{align}
If $J_{0}$ is not a half integer, for $n \in \Z,$ we get 
\begin{equation}
\langle J_{0}, \varphi_{0}|A_{J}|J_{0}, \varphi_{0}\rangle =  0, \quad \langle J_{0}, \varphi_{0}|A_{J^{2}}|J_{0}, \varphi_{0}\rangle = \frac{1}{12}.
\end{equation}
Besides, for $1/2 < \sigma \leq  1$, if there exists $n_{0} \in \Z$ such that $J_{0} \in [n_{0}-\sigma, n_{0} + \sigma]$, then 
\begin{equation}
\langle J_{0}, \varphi_{0}|A_{J}|J_{0}, \varphi_{0}\rangle =  
\frac{1}{\mathcal N^{\sigma}(J_{0})}\frac{n_{0}}{2\sigma} , \quad 
\langle J_{0}, \varphi_{0}|A_{J^{2}}|J_{0}\rangle, \varphi_{0}\rangle =  
\frac{\sigma^{2}}{3} + \frac{1}{\mathcal N^{\sigma}(J_{0})}\frac{n^{2}_{0}}{2\sigma}.
\end{equation}

\item[(iii)]{Angle} 

The lower symbol of the angle operator is given by:
\begin{eqnarray}
\langle J_{0}, \varphi_{0}|A_{\varphi}|J_{0}, \varphi_{0}\rangle &=& 
\pi + 
i\left(1-\frac{1}{2\sigma}\right)\frac{1}{\mathcal N^{\sigma}(J_{0})} \frac{1}{2\sigma}   \crcr
&& \times  \sum_{n\in \Z}
\left[e^{i\varphi_{0}}\chi_{[n-\sigma, n-1 +\sigma]}(J_{0}) - e^{-i\varphi_{0}}\chi_{[n+1-\sigma, n +\sigma]}(J_{0})\right].
\end{eqnarray}
In the lowest limit $\sigma = 1/2$,   we get for  $J_{0}  \, \mathrm{a.e.} \in \R$,  
$\langle J_{0}, \varphi_{0}|A_{\varphi}|J_{0}, \varphi_{0}\rangle = \pi$.

\item[(iv)]{Fourier exponentials} 

The respective lower symbols of the elementary Fourier exponentials are given by:

\begin{equation}
\langle J_{0}, \varphi_{0}|A_{e^{ \pm i \varphi}}|J_{0}, \varphi_{0}\rangle = \left(1-\frac{1}{2\sigma}\right)
\frac{1}{\mathcal N^{\sigma}(J_{0})}\frac{e^{\pm i \varphi_{0}}}{2\sigma}\sum_{n \in \Z}
\chi_{[n+1-\sigma, n+\sigma]}(J_{0}), 
\end{equation}
 

\item[(v)]{Commutator action-angle}
 
The lower symbol of the commutator $[A_{J}, A_{\varphi}]$ is given by:
\begin{align}
\langle J_{0}, \varphi_{0}|[A_{J}, A_{\varphi}]|J_{0}, \varphi_{0}\rangle  
\nonumber =&i\left(1-\frac{1}{2\sigma}\right) \frac{1}{\mathcal N^{\sigma}(J_{0})}\frac{1}{2\sigma} \\
 \times& \sum_{n\in \Z}
\left[e^{i\varphi_{0}}\chi_{[n-\sigma, n-1 +\sigma]}(J_{0}) + e^{-i\varphi_{0}}\chi_{[n+1-\sigma, n +\sigma]}(J_{0})\right]\, . 
\end{align}

\item[(vi)]{Commutator energy-Fourier exponentials} 

The same computation for the commutator $[A_{J^{2}}, A_{e^{\pm i\varphi}}]$ gives 
\begin{align}
\nonumber &\langle J_{0}, \varphi_{0}|[A_{J^{2}}, A_{e^{i\varphi}}]|J_{0}, \varphi_{0}\rangle = \\
&\left(1-\frac{1}{2\sigma}\right)
\frac{e^{i\varphi_{0}}}{2\sigma\mathcal N^{\sigma}(J_{0})}\sum_{n \in \Z}
(2n+1)\chi_{[n+1-\sigma, n+\sigma]}(J_{0})\, ,\\
\nonumber &\langle J_{0}, \varphi_{0}|[A_{J^{2}}, A_{e^{-i\varphi}}]|J_{0}, \varphi_{0}\rangle = \\
&\left(1-\frac{1}{2\sigma}\right)\frac{e^{-i\varphi_{0}}}{2\sigma\mathcal N^{\sigma}(J_{0})}\sum_{n \in \Z}
(1-2n)\chi_{[n-\sigma, n-1+\sigma]}(J_{0})\, .
\end{align}
\end{enumerate}

Note that we can now  use  the normalization function obtained in (\ref{normun}) to
simplify some of the expressions of the above lower symbols as follows:
\begin{equation}
\frac{1}{2\sigma}\sum_{n\in \Z}\chi_{[n+1-\sigma, n +\sigma]}(J_{0}) =\mathcal N^{\sigma}(J_{0})  - \frac{1}{2\sigma} \, , 
\end{equation}
\begin{eqnarray}{\label{lows00}}
\langle J_{0}, \varphi_{0}|A_{e^{ \pm i\varphi}}|J_{0}, \varphi_{0}\rangle 
&=& e^{\pm i\varphi_{0}}\left(1-\frac{1}{2\sigma}\right)\left(1 - \frac{1}{2\sigma \mathcal N^{\sigma}(J_{0})}\right)\, , 
\end{eqnarray}
\begin{equation}{\label{lows01}}
\langle J_{0}, \varphi_{0}|A_{\varphi}|J_{0}, \varphi_{0}\rangle = \pi - 
2\left(1-\frac{1}{2\sigma}\right)\left(1 - \frac{1}{2\sigma \mathcal N^{\sigma}(J_{0})}\right)\sin{\varphi_{0}}\, .
\end{equation}
We note that for   $\varphi_{0} = \pi$,  the lower symbol  
 (\ref{lows01})  is equal to the classical average  $\pi$.

Moreover we get
\begin{equation}
\label{ }
\langle J_{0}, \varphi_{0}|[A_{J}, A_{\varphi}]|J_{0}, \varphi_{0}\rangle 
=2 i\left(1-\frac{1}{2\sigma}\right)\left(1 - \frac{1}{2\sigma \mathcal N^{\sigma}(J_{0})}\right)\cos{\varphi_{0}}.
\end{equation}
Thus  we  recover the canonical commutation rule up to a multiplicative factor. 
Actually, for $J_0$ such that $N^{\sigma}(J_{0}) = 1/(2\sigma)$ the above expression is 0 whereas in the other case  
$N^{\sigma}(J_{0}) = 1/(\sigma)$ it is equal to $ i(1-1/(2\sigma))\cos{\varphi_{0}}$. It is also $0$ for $\varphi_0 =\pi/2$ or $3\pi/2$.

\section{Conclusion}
\label{conclu}
We have reviewed the general procedure of quantization for a given set $X$ of parameters equipped with a measure $\mu$ and studied some relevant 
statistical features by taking into account the interplays between  discrete and continuous probability distributions. The motion of a 
particle on a circle has been studied by considering the cylinder as the corresponding phase space which, in this context, plays the role of  set $X$. 
We have constructed various families of coherent states which are determined by  probability distributions on the cylinder. These distributions are requested   to obey a minimal set of  properties which still leave a large spectrum of possibilities.     The resulting quantization of classical observables has been implemented. The relations between the derived quantum operators together with their respective commutators have been analyzed either directly from the properties of the corresponding operators or through  the behavior of their respective lower symbols. The method has been illustrated in more details with the particular case of uniform distributions on intervals. An interesting feature of our formalism lies in the possibility of applications to models encountered in nanophysics  like circuit QED (see \cite{naciu10} and references therein). Indeed, the question of validity of a precise choice of probability distribution could be experimentally tested in such a context.   Another domain of applications where such tests are possible is the so-called quantum chaos appearing in systems with cylindric phase space, e.g. the kicked pendulum or rotator for which the consistency between  CS and standard quantizations  of the  
Chirikov  standard map \cite{haake-shepelyansky,changshi86} (and references therein) should be fully validated.   
Finally, note that  it should be interesting to deepen  the Euclidean symmetry (Equations (\ref{covCS}) and  (\ref{creucl1})) preserved by CS quantization 
in the spirit of  the  Perelomov's group theoretical methods for building  generalized 
 coherent states \cite{perelomov}.
 All these  questions require the elaboration of appropriate  theoretical framework  which is now under 
investigation and will be in the core of a forthcoming paper.

\section*{Acknowledgments}
JPG expresses his gratitude to  the  
ICMPA-UNESCO Chair and 
University of Abomey-Calavi for their financial support and hospitality, and to the French Ministry of Foreign Affairs for financial support.

\appendix

\section{Statistical aspects of CS quantization}
\label{statasp}

First, the transform $f\mapsto \check{f}$ is built from the nonnegative kernel $\vert \lg x| x'\rg\vert^2$ which is also a family of probability distributions, indexed by $x\in X$, on the set $X$  equipped with the measure $\mathcal{N}(x')\mu(dx')$. Hence, the function $x\mapsto \check{f}(x)$ is the average of $f$ with respect to the latter. Here is encountered a sort of regularization of the original $f$ (depending of course of the topology affected to $X$).

Second, there is also an interplay between two probability distributions \cite{main:ch5:algahel,gazbook09}:
\begin{itemize}
  \item For almost each $x$, a discrete distribution, 
 \begin{equation}
\label{discrprob}
n \mapsto \vert \lg e_n|x\rg\vert^2=  \frac{ \vert \phi_n (x) \vert^2}{{\cal N} (x)} \, .
\end{equation}
Within a quantum physics framework, this probability could be considered  as concerning experiments performed on the system described by the Hilbert space $\mathcal{H}$ within some experimental protocol, say $\mathcal{E}$, in order to measure the  spectral values of a certain self-adjoint operator, a 
\textquotedblleft quantum observable\textquotedblright,  $A$ acting in $\mathcal{H}$ and having the discrete spectral resolution $A = \sum_n a_n |e_n\rg\lg e_n|$.

 \item For each $n$, a ``continuous'' distribution on $(X,\mu)$, 
\begin{equation}
\label{contprob}
X \ni x \mapsto \vert \phi_n (x) \vert^2\,.
\end{equation}
 \end{itemize}
 
 Here,  we observe a Bayesian duality typical of coherent states \cite{main:ch5:algahel}. There are two interpretations: the
resolution of the unity verified by  the   ``coherent'' states $|x\rg$,
introduces a preferred  {\em prior measure}  on the  set  $X$, which is the set of parameters of  the discrete distribution, with this distribution itself playing the role of the
{\em likelihood function}. 
The associated discretely indexed continuous distributions become the related
{\em conditional posterior distribution}.   
   
Hence, a  probabilistic approach to experimental observations concerning $A$ should serve as a  guideline in choosing the set  of the $\phi_n(x)$'s.

We note that the continuous prior distribution will be relevant for the quantization whereas the discrete posterior one characterizes the measurement of the physical spectrum from which is built the ``coherent'' superposition of quantum states $|e_n\rg$.

\section{Normal law coherent states for the motion on the circle}
\label{normalaw}
The functions $\phi_n (x)$ forming the orthonormal system needed to construct coherent states are chosen as Gaussian weighted Fourier exponentials:
\begin{equation}
\label{ficir15}
\phi_n (x) = \left(\frac{1}{2\pi\sigma^2}\right)^{1/4}\,e^{-\frac{1}{4\sigma^2}(J-n)^2 } \,e^{ i n\varphi}\, , \quad n\in \Z \, ,
\end{equation}
where $\sigma > 0$ is a regularization parameter that can be arbitrarily small.  
The coherent states \cite{main:ch5:debgo,main:ch5:kopap,main:ch5:delgo} read as 
\begin{equation}
\label{ccs15}
|x \rg \equiv | J, \varphi \rangle =  \frac{1}{\sqrt{{\mathcal N}^{\sigma} (J)}}\,  \left(\frac{1}{2\pi\sigma^2}\right)^{1/4}
 \sum_{n \in \Z} e^{-\frac{1}{4\sigma^2}(J-n)^2 } \,e^{- i n\varphi} | e_n\rangle\, ,
\end{equation}
where the states $| e_n\rangle$'s, in one-to-one correspondence with the $\phi_n$'s, form an orthonormal basis of some separable Hilbert space 
$\mathcal{H}$.
For instance, they can  be considered as Fourier exponentials
$e^{i n\theta}$ forming the orthonormal basis of the Hilbert space $L^2(S^1,d\theta/2\pi) \cong
\mathcal{H}$. They would be the \emph{spatial or angular modes} in this representation. In this representation, the coherent states read as the following Fourier series:
\begin{equation}
\label{csfs}
\zeta_{J,\varphi} (\theta) = \frac{1}{\sqrt{{\mathcal N}^{\sigma} (J)}}\,  \left(\frac{1}{2\pi\sigma^2}\right)^{1/4} \sum_{n \in \Z} e^{-\frac{1}{4\sigma^2}(J-n)^2 } \,e^{ i n(\theta -\varphi)}\,. 
\end{equation}
The normalization factor is  a periodic train of normalized Gaussians which can be written as an elliptic theta function \cite{magnus66}:
 \begin{equation}
\label{norci15}
 \mathcal{N}^{\sigma}(J) = \sqrt{\frac{1}{2\pi\sigma^2}}\sum_{n \in \Z} e^{-\frac{1}{2\sigma^2} (J-n)^2} = \vartheta_3(J,2\pi i\sigma^2)\underset{\mbox{Poisson}}{=}  \sum_{n \in \Z} e^{2\pi i nJ}\, e^{-2\sigma^2\pi^2 n^2} \, .
\end{equation}
Its asymptoptic behavior at small and large values of the parameter $\sigma$ is given by 
\begin{align}
\label{sigma0}
&\lim_{\sigma \to 0}\mathcal{N}^{\sigma}(J) = \sum_{n\in \Z} \delta(J-n)\quad \mbox{(Dirac comb)}\,,  \\
\label{sigmainf} & \lim_{\sigma \to \infty}\mathcal{N}^{\sigma}(J) = 1\, .
\end{align}
We also note that $\lim_{\sigma \to 0} \sqrt{2\pi\sigma^2}\mathcal{N}^{\sigma}(J) = 1$ if $J\in \Z$ and $= 0$ otherwise.

By construction, the states (\ref{ccs15}) are normalized and resolve the identity in the Hilbert space $\mathcal{H}$:
\begin{equation}
\label{ccsresid}
\int_{-\infty}^{+\infty}dJ\int_0^{2\pi}\frac{d\varphi}{2\pi} \, \mathcal{N}^{\sigma}(J)\, |J,\varphi\rg \lg J,\varphi| = \I_{\mathcal{H}}\, .
\end{equation}
They overlap as
\begin{align}
\label{overlapgcs1}
\lg x|x'\rg &=  \frac{e^{-\frac{1}{8\sigma^2}(J-J')^2}}{\sqrt{ 2\pi \sigma^2\, \mathcal{N}^{\sigma}(J)\, \mathcal{N}^{\sigma}(J')}}\sum_{n\in \Z} e^{-\frac{1}{2\sigma^2}(\frac{J+J'}{2}- n)^2}\, e^{i  n(\varphi-\varphi') }\\
\label{overlapgcs2}&\underset{\mbox{Poisson}}{=} 
 \frac{e^{-\frac{1}{8\sigma^2}(J-J')^2}\,e^{i\frac{J+J'}{2}(\varphi -\varphi')}}{\sqrt{ \mathcal{N}^{\sigma}(J)\, 
\mathcal{N}^{\sigma}(J')}}\sum_{n\in \Z} e^{-\frac{\sigma^2}{2}(\varphi-\varphi'-2\pi n)^2}\, e^{-i \pi n(J+J') }\,.
\end{align}
These expressions stand for the representation of the coherent state $|x'\rg$ as a function of  $x=(J,\varphi)$. It is interesting to explore the two possible limits of the Gaussian width:
\begin{align}
\label{sigma01}
 &\lim_{\sigma \to 0}  \lg x|x'\rg =   \left\lbrace\begin{array}{cc}
    0 &  \mbox{if} \quad J \notin  \Z \ \mbox{or} \  J' \notin  \Z\\
     \delta_{JJ'} \, e^{iJ(\varphi-\varphi')}  &    \mbox{if} \quad J \in \Z
\end{array}\right.  \, ,  \\
\label{sigmainf1}  &\lim_{\sigma \to \infty} \lg x|x'\rg     =   \left\lbrace\begin{array}{cc}
    0 & \mbox{if} \quad \varphi-\varphi' \notin 2\pi \Z  \\
    1  &  \mbox{if} \quad \varphi-\varphi' \in 2\pi \Z
\end{array}\right. \, 
\end{align}
where $\delta_{JJ'}$ is the Kronecker symbol, i.e. $=0$ if $J\neq J'$ and $=1$ if $J=J'$. 
Therefore, from (\ref{sigma01}),   the coherent states  tend to be orthogonal at small $\sigma$ if $J\notin\Z$ or if $J\neq J'$ 
whatever the value of the difference $\varphi-\varphi'$. On the other hand, from (\ref{sigmainf1}),  
the coherent states tend to become orthogonal at large $\sigma$  if $\varphi-\varphi' \notin 2\pi \Z$, 
whatever the value of the difference $J-J'$. We have here an interesting duality in semi-classical aspects of these states, 
the term ``semi-classical'' being used for both limits of the parameter $\sigma$. We will come to this important point at the end of 
Section \ref{quantcs}.

\end{document}